\documentclass[prb,amsmath,onecolumn]{revtex4-1}
\usepackage{graphicx}
\usepackage{amssymb}
\usepackage{color}

\usepackage{float}

\begin{document}

\title{Free Triplets versus Bound Triplet-Triplet Biexciton in Intramolecular Singlet Fission Materials:
Structure-Property Correlations}
\author{Souratosh Khan}
\affiliation{School of Information, University of Arizona
Tucson, AZ 85721}
\author{Sumit Mazumdar$^*$}
\affiliation{Department of Physics, University of Arizona}
\affiliation{Department of Chemistry and Biochemistry, University of Arizona}
\affiliation{College of Optical Sciences, University of Arizona}
\email{mazumdar@email.arizona.edu}
\date{\today}

\begin{abstract}
Recent advances in singlet-fission research make it imperative that structure-property correlations that
determine optical signatures of the triplet-triplet spin biexciton
as well as its binding energy be understood precisely.
We report many-body calculations
of excited state absorptions from the triplet exciton and the triplet-triplet biexciton from two transversally linked dimers of pentacene
derivatives. Comparison of experiment against theory leads to new interpretations of experiments performed earlier.
We show that in
the para-linked isomer the triplet-triplet does not dissociate to free triplets through the duration of the measurements. 
In contrast, even as calculated and experimental transient absorptions agree in the meta-isomer, the experimental observations here 
are more difficult to interpret, indicating the strong role structural variations can play in determining the rate and yield of
free triplets. We also report many-body calculations of the spin gap, the energy difference between the
spin quintet versus spin singlet triplet-triplet, as well as the
binding energy of the spin singlet triplet-triplet, defined as the energy difference
between two free triplets and the bound biexciton. The spin gap and the binding energy of the spin singlet triplet-triplet 
are different quantities in all but coupled two-level systems.  
The experimental behavior in the transversally linked dimers as well as previously studied longitudinally linked dimers agree with the 
trends that would be predicted from the computed biexciton binding energies.
\end{abstract}

\maketitle

\section{Introduction}

Singlet fission (SF) is a photophysical process that
involves the generation of two spin-triplet excitons (T$_1$)
from a single optically accessible singlet exciton (S$_1$) in an organic $\pi$-conjugated molecule. 
As the process generates four charge-carriers per absorbed photon 
it is being intensively investigated \cite{Michl13b,Smith13a,Rao17a,Casanova18a} as a possible means to overcome
the Shockley-Quessier limit \cite{Shockley61a} for the efficiencies of single junction organic solar cells. SF requires excitations
across multiple chromophore molecules, and interest has shifted in recent years from intermolecular to intramolecular SF (ISF) in
covalently linked chromophore molecules \cite{Sanders15a,Zirzlmeier15a,Lukman15a,Busby15a,Fuemmeler16a,Sakuma16a,Sanders16a,Sanders16c,Korovina16a,Liu15b,Margulies16a,Korovina18a}. Very recently, 
experimental research has been extended to oligomers consisting of up to five acene monomers, which are not all same \cite{Pun19a}.

SF is a spin-allowed multistep process in which the S$_0$S$_1$ state (here S$_0$ is the monomer ground state) first relaxes to a bound triplet-triplet biexciton $^1$(TT)$_1$ 
that is overall spin singlet (here the superscript and subscript refer to the spin multiplicity and the quantum 
number of the state within the triplet-triplet space, respectively). We note that the $^1$(TT)$_1$ is in the even spatial parity spin singlet subspace and can occur below S$_1$. 
Our nomenclature allows a clear distinction between one- versus two-photon spin singlet states. 
$^1$(TT)$_1$, nominally a double excitation within molecular orbital (MO) theory \cite{Hudson82a}, is often degenerate with or even lower in energy than S$_0$S$_1$ in
due to strong Hubbard repulsion among $\pi$-electrons occupying the same $p_z$ orbital \cite{Ramasesha84b, Ramasesha84c, Tavan87a}. SF should be considered complete 
only when the $^1$(TT)$_1$ further dissociates into a pair of free triplets T$_1$. In ISF, the assumption has often been that the $^1$(TT)$_1$ is weakly bound
and triplet energy transfer will occur from the photoexcited dimer to a neighboring dimer in its ground state, leading to two free triplets.

Since $^1$(TT)$_1$ and T$_1$ are both optically inaccessible from the ground state, they
are identified from ultrafast excited state spectroscopy. One key question in SF is then whether or not there exist experimental optical signatures
of the bound $^1$(TT)$_1$ biexciton that are distinct from those of T$_1$. Identification of unique optical signatures of $^1$(TT)$_1$ is essential for 
the determination of its lifetime.
Further, the dissociation efficiency of $^1$(TT)$_1$ depends on its
binding energy E$_b$, defined as the energy difference between the two free triplets and the triplet-triplet \cite{Zimmerman18a,Musser19a}, 2$\times$E(T$_1)-$E($^1$(TT)$_1$). 
Structural features that determine E$_b$ are also of strong interest.
Determining these have acquired urgency in recent years with the discovery that
the dissociation of $^1$(TT)$_1$ into two free T$_1$ takes much longer than
what was believed until recently. Instead of hundreds of femtoseconds (fs) \cite{Wilson13a, Rao17a}, 
the completion of SF can take upto nanoseconds (ns) \cite{Musser19a,Yong17a,Stern17a,Weiss16a,Tayebjee17a,Basel17a,Trinh17a, Miyata19a}. Thus the dissociation of $^1$(TT)$_1$, and not the
internal conversion of S$_0$S$_1$ to $^1$(TT)$_1$, may be the rate determining step in SF. Concurrent theoretical
work on crystals of pentacene \cite{Khan17c}, covalently linked homodimers bipentacenes BPn \cite{Khan17b} and pentacene-tetracene heterodimers PTn \cite{Khan18a} have shown that ultrafast excited state
absorptions (ESAs) in the visible range of the electromagnetic spectrum, previously ascribed to T$_1$, are from the bound $^1$(TT)$_1$, whose intramonomer
excitations overlap in the visible with those of T$_1$. 
Many-body calculations for BPn and PTn predicted {\it additional} $^1$(TT)$_1$ ESAs in the near infrared (NIR) and short-wave IR (SWIR)  
that are absent in T$_1$ spectra. These IR absorptions have subsequently been detected in BPn and PTn \cite{Trinh17a, Miyata19a}, as well as in oligomers \cite{Pun19a}. 
\begin{figure}
\includegraphics[width=6in]{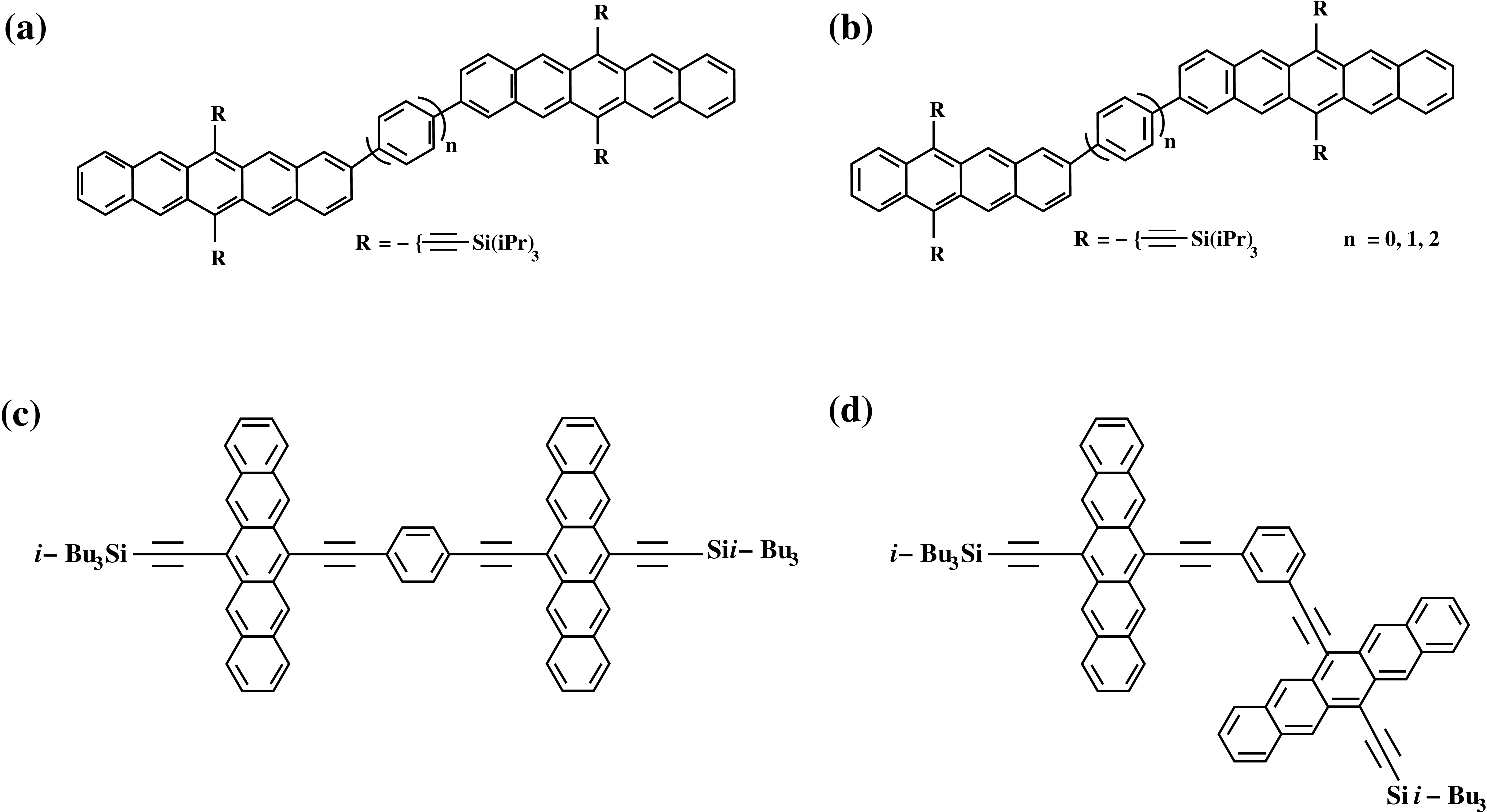}
\caption{
\it TIPS-acene dimers linked via phenylene spacer groups: (a) Bipentacenes BPn, (b) the asymmetric pentacene-tetracene dimers PTn, (c) and (d) para- and meta-(bisethynylpentacenyl)
benzene dimers, $p-2$ and $m-2$. Computational results on the photophysics of BPn and PTn have been discussed previously \cite{Khan17b,Khan18a}. Present work focuses on
$p-2$ and $m-2$, but comparisons are made with the earlier results on BPn and PTn to point out similarities and differences, as appropriate.}  
\label{ZirzlBP0}
\end{figure}

BPn and PTn consist of acene monomers linked {\it longitudinally} through $n=0-3$ phenylene linkers
(2-2$^\prime$ links, see Figs.~1(a) and (b)). The limited geometries investigated theoretically so far raise new questions crucial for understanding the mechanism of ISF.
First, are the ESAs in the IR from $^1$(TT)$_1$ expected in molecular dimers irrespective of topology, or are they unique to specific structural features (such as 2-2$^\prime$ links) ? 
Second, what is the relationship, if any, between these absorptions and E$_b$? 
Finally, since ultrafast measurements in the IR are difficult, can the   
{\it qualitative trends} in E$_b$ be guessed from other measurements? 

To resolve the above questions we have investigated theoretically dimers of TIPS-pentacene (TIPS= triisopropylsilyl) that are structurally maximally different from BPn and PTn.
Not only are the C-C triple bonds now involved in the inter-monomer conjugation (unlike in the 2-2$^\prime$ linked BP1), the monomers 
are linked {\it transversally} through a phenylene linker (6-6$^\prime$ link, see Figs.~1(c) and (d)), 
as opposed to longitudinally. We investigate theoretically the experimental claim
of completed SF in both, that was based on monitoring transient absorptions in the visible alone \cite{Zirzlmeier15a}.
We adopt the same short-hand nomenclatures for the molecules as in the original paper \cite{Zirzlmeier15a}, $p-2$ and $m-2$, to label dimers 
6-6$^\prime$ linked through para- 
and meta- linkages via the phenylene. 
We also examine the theoretical claim \cite{Abraham17a}
that in $m-2$ the spin quintet $^5$(TT)$_1$ is lower in energy than the singlet $^1$(TT)$_1$, and 
that E$_b$ is negative (which would imply spontaneous direct 
decay from S$_0$S$_1$ to 2 $\times$ (T$_1$)).
We report here accurate many-body computational results of ESAs from T$_1$, and importantly, $^1$(TT)$_1$ in $p-2$ and $m-2$, for comparison to experiments.
In addition, we report calculations of the spin gap $\Delta_S$ = E($^5$(TT)$_1)-$ E($^1$(TT)$_1$),
and E$_b$, in $p-2$ and $m-2$, as well as linear polyenes and BP1 to arrive at generic qualitative answers to the questions we have posed above. 
We recognize that $\Delta_S$ and E$_b$ are small and the uncertainties in our computationally obtained quantities are nonnegligible.
We are however confident that the ranges and the overall trend for the quantities computed within our many-body approach are accurate and more importantly, 
that the predicted structure-property trends (2-2$^\prime$ versus 6-6$^\prime$, and para versus meta links) are correct.  

\section{Theoretical Model, Parametrization and Computational Methods}

We consider the $\pi$-electron only Pariser-Parr-Pople (PPP) Hamiltonian \cite{Pariser53a,Pople53a}.
\label{PPP_Ham}
\begin{eqnarray}
H=\sum_{\langle ij \rangle,\sigma}t_{ij}(c_{i\sigma}^{\dagger}c_{j\sigma}+c_{j\sigma}^\dagger c_{i\sigma}) \\ 
\nonumber
+ U\sum_{i}n_{i\uparrow}n_{i\downarrow} + \sum_{i<j} V_{ij} (n_i-1)(n_j-1) \\
\nonumber
\end{eqnarray}
where $c^{\dagger}_{i\sigma}$ creates an electron with spin $\sigma$ on the $p_z$ orbital of carbon (C) atom
$i$, $n_{i\sigma} = \sum c^{\dagger}_{i\sigma} c_{i\sigma}$ is the number
of electrons with spin $\sigma$ on atom $i$, and $n_i=\sum_{\sigma}n_{i\sigma}$ is the total number of electrons on the atom.
We retain electronic hoppings $t_{ij}$ only between nearest neighbors $i$ and $j$.
$U$ is the Coulomb repulsion between two electrons occupying the $p_z$ orbital of the same C-atom, and
$V_{ij}$ is long range Coulomb interaction. The average bond lengths within an acene unit are different for the peripheral (1.40 $\mathring{\textrm{A}}$)
and internal (1.46 $\mathring{\textrm{A}}$) C-C bonds \cite{Khan18a}.
Based on a widely used bond length-hopping integral relatonship  \cite{Ducasse82a}
we have chosen intra-acene peripheral (internal)
hopping integrals $t_{ij}$ as $-$2.4 ($-$2.2) eV.
For the C-C triple bonds we have chosen $t_{ij}=-3.0$ eV \cite{Ducasse82a}.
It is known that $p-2$ is planar and $m-2$ is nearly planar \cite{Zirzlmeier15a}; we have chosen planar geometries for both
and therefore interunit C-C hopping integrals $-$2.2 eV \cite{Ducasse82a} between the TIPS-pentacene monomers and the phenylene linker.
We use the screened Ohno parameterization for the long range Coulomb repulsion, $V_{ij}=U/\kappa\sqrt{1+0.6117 R_{ij}^2}$,
where $R_{ij}$ is the distance in $\mathring{\textrm{A}}$
between C-atoms $i$ and $j$ and $\kappa$ is an effective dielectric constant \cite{Chandross97a}. 
The parameters $U$ and $\kappa$ were chosen from comparisons to known
monomer TIPS-pentacene energies. Monomer E(S$_1$) is reproduced best with $U=6.7$ eV, $\kappa = 1.0$. 
However, the dipole-allowed triplet excitation energy, E(T$_3$)$-$E(T$_1$), of interest here, is best reproduced with $U=7.7$ eV, $\kappa = 1.3$ 
(see Table S1, Supporting Information; (E(T$_1$) is almost the same for both parameter sets). 
The justification for using $U$ smaller than that within the ``standard'' Ohno parameters \cite{Ohno64a} and $\kappa \neq 1$ come from extensive
fittings of wavelength dependent spectra in $\pi$-conjugated polymers \cite{Chandross97a} as well as polyacenes \cite{Sony07a}, with multiple $U$
and $\kappa$.

We report results for both sets of close lying parameters. Our inclusion of both allows obtaining
an accurate range for the calculated $\Delta_S$ and E$_b$, while the dominant exciton basis wavefunction components of interest (see below) are
the same for the two parameters.
\begin{figure*}
\includegraphics[width=\textwidth]{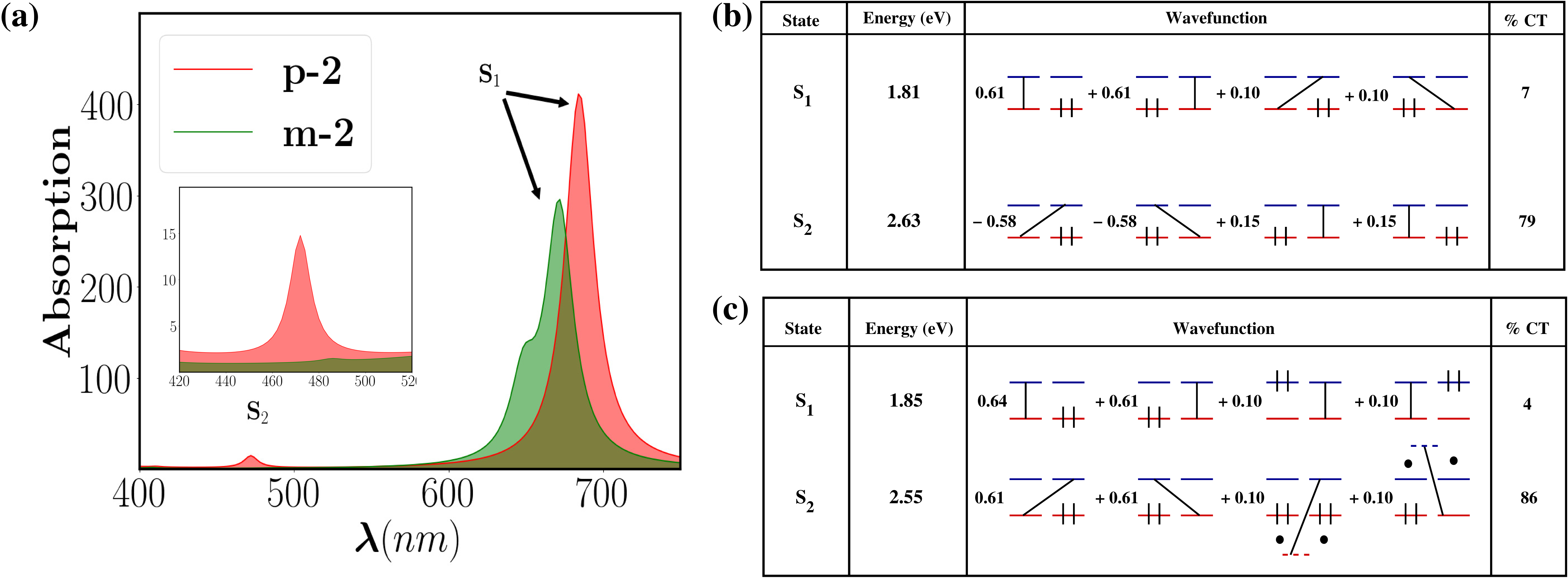}
\caption {\it (a) Calculated ground state absorption spectra for $p-2$ and $m-2$ dimers with U = 6.7 eV
and $\kappa=1.0$, to be compared against Figs. S15(b) and S14, respectively, in the Supporting Information of
reference \onlinecite{Zirzlmeier15a}. 
The shoulder on the absorption band of $m-2$ is due to absorption to a state that becomes weakly allowed due to the absence of
inversion symmetry. 
The inset shows the relative intensities of the CT absorptions. 
(b) and (c) Dominant exciton basis
contributions to S$_1$ and S$_2$ wavefunctions, in $p-2$ and $m-2$, respectively. The last column gives the percentage CT contribution to the wavefunction. 
}
\label{ZirzlBP1}
\end{figure*}

The PPP Hamiltonian allows rigorous many-body calculations of the energies of and ESAs from the $^1$(TT)$_1$ that are not possible for large molecules within first principles approaches.
Accurate determinaton of just the energy of this two electron--two hole (2e-2h) excitation requires including configuration interaction (CI) with
at least 4e-4h excitations from the Hartree-Fock (HF) ground state \cite{Tavan79a}. This continues to be difficult within first principles approaches \cite{Casanova18a,Zimmerman18a} 
for molecules with more than about $\pi$-electrons
and certainly for the present case with 58 $\pi$-electrons. Calculating ESAs from $^1$(TT)$_1$, or
$\Delta_S$ and E$_b$, 
make the requirements on the theory even more stringent. We use here a modified version of the 
multiple reference singles and doubles CI (MRSDCI) approach, that was originally developed to include the dominant 1e-1h, 2e-2h and 4e-4h excitations that best describe
any targeted excited state \cite{Tavan87a}, including $^1$(TT)$_1$. We have modified the original technique in order to obtain simultaneously the ESA spectrum, by 
including among the reference configurations
not only the minimal basis required to obtain the targeted state, but also the configurations that are dipole-coupled to the fundamental reference configurations
(see Section B, Supporting Information.) Each targeted state (S$_1$, T$_1$, $^1$(TT)$_1$ and  $^5$(TT)$_1$) and the final states of the ESAs from it are thus obtained by solving the same MRSDCI
Hamiltonian matrix. In every case our Hamiltonian matrices have dimensions of several million (see Tables S2 and S3, Supporting Information).

Our calculations are done using a localized exciton basis that allows pictorial representations of eigenstates.
\cite{Khan17b,Khan17c,Khan18a}. The Hamiltonian (Eq.~1) is written as $H=H_{intra}+H_{inter}$, where $H_{intra}$ consists of purely intramolecular 
terms within Eq.~1 and $H_{inter}$ consists of the remaining intermolecular terms. HF MOs that are solutions of $H_{intra}$ are obtained in the first step of the calculations.
MRSDCI diagonalization of $H_{intra}+H_{inter}$ then yields eigenstates of the complete Hamiltonian as superpositions of 
many-electron configurations in which these HF MOs are occupied by electrons in
all possible manner, including upto 4e-4h excitations. A thorough discussion of the application of the exciton basis that illustrates all
the finer points can be found in reference \cite{Chandross99a}, which reported exact PPP calculations for trans-decapentaene, with H$_{intra}$ describing
individual ethylenic units.
The advantage of this description is that not only excitations can be classified as predominantly intra- versus intermonomer,
final states of dipole-allowed optical excitations from any initial state can be anticipated from the diagrammatic representation of the initial state. The latter constitutes a
strong check on the numerical calculations.


\section{Results and Analysis}

\begin{table*}[t]
\small
\centering
\caption{\textbf{Calculated energies of S$_1$, T$_1$, $^1$(TT)$_1$, $^5$(TT)$_1$ (in eV) in polyenes and acene dimers for U = 6.7 eV, $\kappa$ = 1.0 
(outside brackets) 
and U = 7.7 eV, $\kappa$ = 1.3 (inside brackets).}}
\setlength{\tabcolsep}{10pt}
\begin{tabular}{l  c c c c }
\hline\hline
   & E(S$_1$) & E(T$_1$)  & E($^1$(TT)$_1$)  & E($^5$(TT)$_1$) \\[0.5ex]
\hline

trans-octatetraene & 3.56 (4.09)  & 2.06 (1.85) & 3.91 (3.69) & 6.27 (5.55) \\[0.5ex]
trans-dodecahexaene & 2.99 (3.5) & 1.75 (1.63) & 3.18 (3.03) & 4.88 (4.37) \\[0.5ex]
BP1 & 1.91 (2.09) & 1.04 (0.96) & 1.95 (1.75) & 1.96 (1.76) \\[0.5ex]
$p-2$ & 1.81 (1.99) & 0.99 (0.91) & 1.90 (1.71) & 1.95 (1.76) \\[0.5ex]

$m-2$ & 1.85 (2.02) & 0.972 (0.879) & 1.935 (1.736) & 1.937 (1.741) \\[0.5ex]

\hline
\end{tabular}
\label{results}
\end{table*}

In Table 1 we have given the energies of S$_1$, T$_1$ and $^1$(TT)$_1$ and $^5$(TT)$_1$ for both $p-2$ and $m-2$ for both sets of parameters. We first discuss
the singlet, triplet and the $^1$(TT)$_1$ and then follow up with discussions of $^5$(TT)$_1$, and our calculated $\Delta_S$ and E$_b$.
We have included in the Table the same quantities for two linear polyenes as well as BP1, for comparison and understanding of 
structure-dependence of all quantities.
Our calculated $^1$(TT)$_1$ is either nearly degenerate
with or lower in energy than S$_1$ for both $p-2$ and $m-2$, in agreement with experiments
\cite{Zirzlmeier15a}. 
In Fig.~2(a) we have shown the calculated ground state absorptions for $p-2$ and $m-2$, while Figs.~2(b) and 2(c) give the corresponding wavefunctions. The weak CT
absorptions found theoretically are seen at $\sim$ 450 nm experimentally (see Figs.~S14 and S15 of Supplemental Information of Reference \onlinecite{Zirzlmeier15a}). 
Similar (but stronger) CT absorptions are seen also in BPn and PTn, experimentally \cite{Sanders15a,Sanders16c} and within  our many-body computations \cite{Khan17b,Khan18a}.  
There is a subtle difference between the CT contributions to $p-2$ and $m-2$. 
S$_1$ in $p-2$ is moderately strongly coupled to the lowest energy CT state (see wavefunction in Fig.~2(b)). In contrast,
the 4\% CT contribution to $m-2$ comes almost entirely from the higher energy CT diagrams that also contribute to S$_2$ in $m-2$
(see wavefunction in Fig.~2(c)). Since the relative weights of the higher energy configurations in S$_2$ are very small (see Fig.~2(c))
the absorption to S$_2$ in $m-2$ is much weaker, which in turn is a signature of the 
weaker coupling between the TIPS monomers in this compound, as is ascertained also from other calculations reported below. 

Calculations in the spin triplet subspace further confirm the difference in the intermonomer couplings between $p-2$ and $m-2$. T$_1$ in $p-2$ is a superposition of triplet
Frenkel excitons in the monomers (see Fig.~S2(a) in Supporting Information), as is true also for BPn and PTn. There exists an excited triplet CT state T$_2$ that 
is nearly degenerate with S$_2$ (see Fig.~S2(a), Supporting Information).
The weak coupling between the monomers in $m-2$, suggested already from the singlet wavefunctions in Fig.~2(c), 
leads to extreme localization and triplet states that are unique to $m-2$ among all ISF dimers we have studied so far: 
instead of a T$_1$ that is a superposition of Frenkel excitons, triplet eigenstates here occur as {\it distinct degenerate pairs of excitons localized on individual
monomers.} (see Fig.~S2(b) in Supplementary Information).
In Fig.~3(a) 
we have shown the calculated ESAs from the T$_1$ exciton in $p-2$ (red) and $m-2$ (green).
The absorptions in the 550$-$600 nm region, common to both
$p-2$ and $m-2$, are due to intramonomer molecular excitations. The absorption at $\sim$ 700 nm in $p-2$ is to T$_2$, which is of CT character and occurs also in 
BP1 and PT1 \cite{Khan18a}. 
Transient absorptions from T$_1$ are then predicted to be different in $p-2$ and $m-2$.
\begin{figure}
\includegraphics[width=6.in]{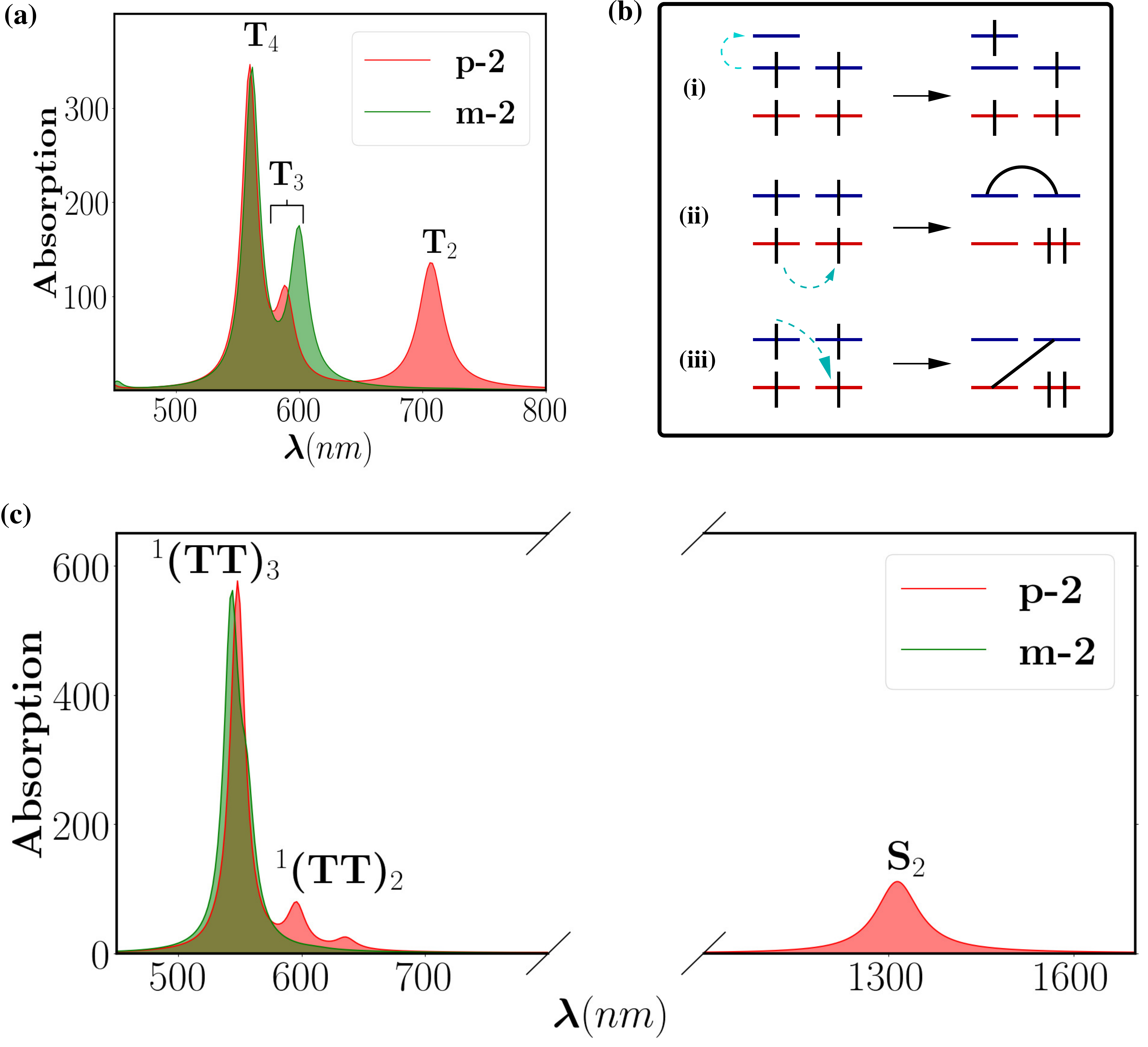}
\caption {\it (a) Calculated triplet ESA spectra in $p-2$ (red) and $m-2$ (green), with $U = 7.7$ eV and $\kappa=1.3$. Absorption near 700 nm in
$p-2$ is currently a prediction. Absorption in this region is absent in $m-2$. The initial T$_1$ and the final states reached in ESA are given in the Supporting Information.
(b) Schematics of the expected optical transitions from the bound $^1$(TT)$_1$. In addition to the monomeric absorption (i) that would overlap with T$_1$ absorption, 
two additional CT absorptions (ii) and (iii)
are expected. The final state in (iii) is the state S$_2$ of Fig.~2. (c) Calculated ESA spectra of $^1$(TT)$_1$ in $p-2$ (red) and $m-2$ (green),  
with $U=7.7$ eV and $\kappa=1.3$ eV. The absorption to T$_2$ in the red to near IR in the free triplet of $p-2$ 
is absent in the corresponding bound triplet-triplet. Correspondingly, the absorption 
in the IR, at 1300-1400 nm in the bound triplet-triplet of $p-2$, is absent in the free triplet absorption from the same compound.} 
\end{figure}

In Fig.~3(b) we show schematically why
{\it two additional absorptions from $^1$(TT)$_1$, beyond the intramolecular excitation (i) that overlap with T$_1$ intramolecular absorptions} are expected
for intermediate to strong intermonomer coupling.
Determining computationally the higher energy CT absorption (ii), to the final state referred to as $^1$(TT)$_2$ hereafter,
requires the retention of both a large number of monomer MOs \cite{Khan17b} as well as a very large many-electron basis \cite{Tavan87a}, neither of which are possible outside the PPP approach. 
Experimentally, the low energy CT absorption (iii) is more relevant,
as this will occur far from the intramolecular absorptions.
From the exciton basis wavefunctions in Fig~2(b) and the schematic in Fig.~3(b)(iii) we predict   
S$_2$ to be the final state of this transient absorption.  
Assuming $^1$(TT)$_1$ to be nearly at 2$\times$E(T$_1$) or quasidegenerate with S$_1$, it then becomes possible to estimate the approximate energy of the 
long wavelength transient absorption of
 $^1$(TT)$_1$ from physical arguments alone, viz., it should be close to, though not exactly, E(S$_2$)$-$E(S$_1$). 

In Fig.~3(c)
we have given the calculated MRSDCI ESA spectra of $^1$(TT)$_1$ for both $p-2$ and $m-2$. The CT absorptions to $^1$(TT)$_2$ and S$_2$ for $p-2$
are clearly indicated.
As expected from the weak intermonomer coupling in $m-2$, seen already from the calculated ground state absorption and the triplet ESA, 
there is negligible CT contribution to $^1$(TT)$_1$ wavefunction here (see Fig.~S3 Supporting Information), leading to 
vanishing strength of the CT absorption in the IR for $m-2$ in Fig.~3(c)

We are now in a position to compare the calculated transient absorptions to the experimental ones in Reference \onlinecite{Zirzlmeier15a}.
In what follows we refer to the experimental figures in the
Supporting Information of Reference \onlinecite{Zirzlmeier15a}, {\it focusing on 
the false color spectra in Figs~S20(b) and (c).} The experimental transient absorption in Fig.~S20(b) for $m-2$ is very narrow and limited to the visible region, 
in excellent agreement with the calculated ESA spectra for $m-2$ in Figs.~3(a) as well as
Fig.~3(c). In contrast, additional absorption extending into the IR (1.2 - 1.4 eV) is clearly seen in the false color spectrum in Fig.~S20(c) for $p-2$, 
also in excellent agreement with our calculated $^1$(TT)$_1$ spectrum for $p-2$ in Fig.~3(c) (weak quantitative deviations between the calculated and experimental ESA
energies are to be expected within the difficult many-body calculations). Furthermore, 
the considerably broader experimental transient absorption in $p-2$ in the visible (see false color spectrum in Fig.~S20(c) in Reference \onlinecite{Zirzlmeier15a}), 
is in agreement with the calculated ESA spectrum for $p-2$ Fig.~3(c), where contribution from absorption to the high energy CT state $^1$(TT)$_2$ occurs.
We also draw attention to the 1.8 eV (700 nm) region where T$_1$ should absorb, but the experimental photoinduced absorption is very weak.
Based on the persistence of the transient absorption in the IR through the duration of the experiment \cite{Zirzlmeier15a}, we conclude that the lifetime 
of the bound $^1$(TT)$_1$ 
in $p-2$ is far longer than what had been assumed before, and dissociation to free triplets does not occur here.

\begin{table*}[t]
\small
\centering
\caption{\textbf{Calculated spin gap $\Delta_S$ and binding energy E$_b$ of $^1$(TT)$_1$ in polyenes and acene dimers for U = 6.7 eV, $\kappa$ = 1.0  (outside brackets) 
and U = 7.7 eV, $\kappa$ = 1.3 (inside brackets).}}
\setlength{\tabcolsep}{10pt}
\begin{tabular}{l  c c }
\hline\hline
   & $\Delta_s$ & E$_b$\\[0.5ex]
\hline

trans-octatetraene &  1.86 (2.36) & 0.21 (0.01) \\[0.5ex]
trans-dodecahexaene &  1.34 (1.70) & 0.32 (0.23) \\[0.5ex]
BP1 & 0.01 (0.02) & 0.13 (0.17) \\[0.5ex]
$p-2$ & 0.05 (0.05) & 0.08 (0.11) \\[0.5ex]

$m-2$ & 0.005 (0.002) & 0.009 (0.021) \\[0.5ex]

\hline
\end{tabular}
\label{results1}
\end{table*}

\begin{figure}[H]
\centering
\includegraphics[width=3.5in]{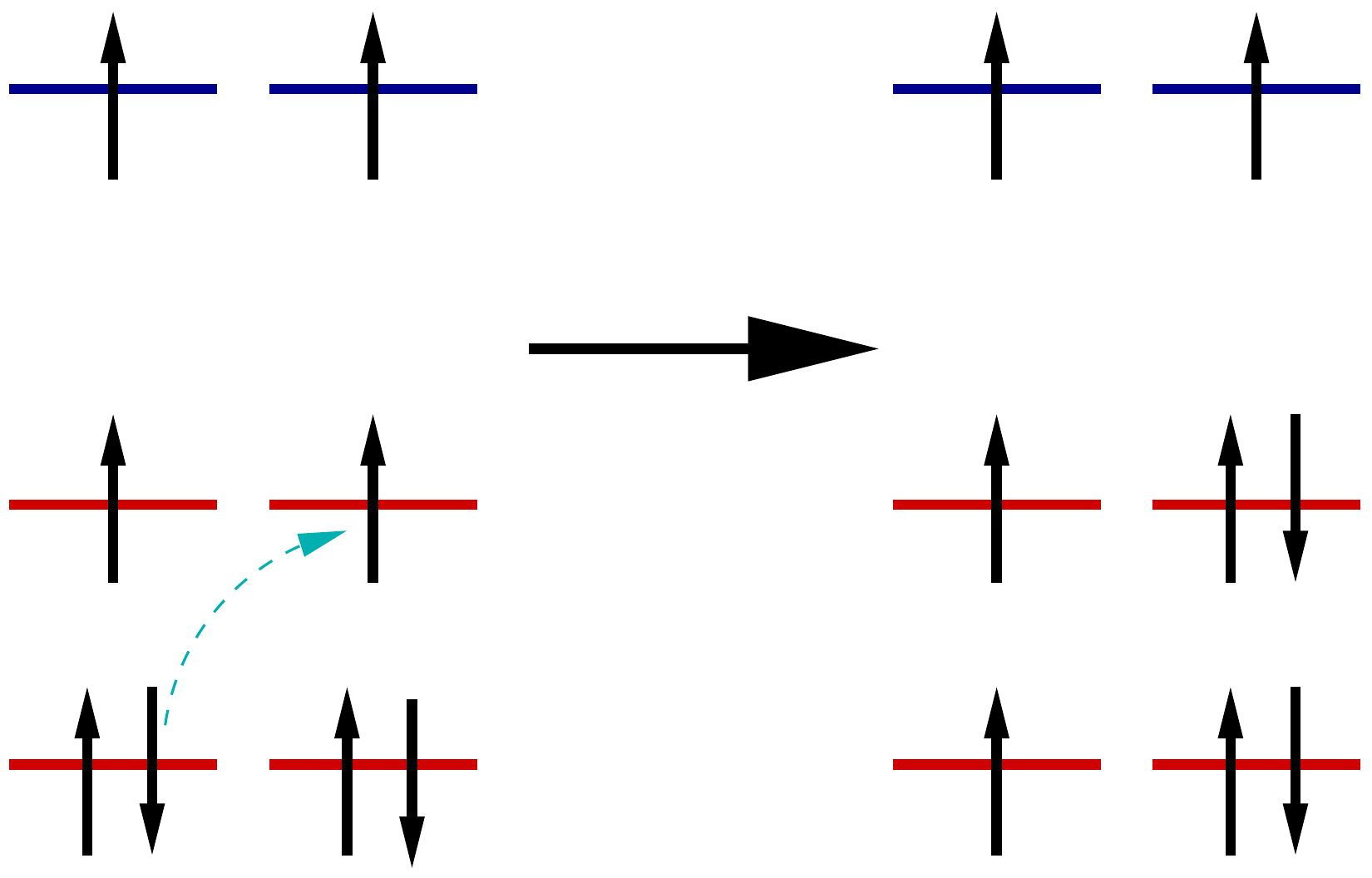}
\label{quintet2}
\caption{Schematic of the CT that will occur in the bound $^5$(TT)$_1$ to lower its energy relative to a pair of free, uncorrelated triplets.
}
\end{figure}

In Table 2 we have given our computed $\Delta_S$ and E$_b$ for all the compounds in Table 1, also for both setes of parameters. There is no correlation between $\Delta_S$ and E$_b$ in the polyenes, which are included for comparison only.
The very large $\Delta_S$ and its decrease with increasing length are both anticipated from the different dominant MO occupancies (see Fig.~S4 in Supporting Information) in 
$^5$(TT)$_1$ versus $^1$(TT)$_1$. In contrast to $\Delta_S$, which decreases with length, the calculated E$_b$ increases with length in this regime, 
which is counterintuitive. This is a finite-size effect. The increase here is because
in the shortest polyenes the two individual triplets in $^1$(TT)$_1$ are strongly overlapping (see schematics in Fig.~S5 of Supporting Information).
While T$_1$ can have optimal length (or close to it) even in short polyenes, the triplets in $^1$(TT)$_1$ overlap and the $^1$(TT)$_1$ is artificially confined,
the combined effect of which is to raise the energy of the biexciton relative to the free triplets \cite{Guo95a}, and to lower E$_b$.
Hence E$_b$ here increases with polyene length until the polyene reaches an optimal length where the triplet overlap is optimal and is decided by the spin-spin coupling 
alone, beyond which E$_b$ should decrease monotonically. 
The situation is different in the acene dimers, where the triplets in $^1$(TT)$_1$ and $^5$(TT)$_1$ occupy different monomers (see Fig.~S3, Supporting Information) 
and are hence nonoverlapping. 
The orbital occupancies in the exciton basis
are thus the same for the spin singlet and spin quintet triplet-triplet.
$\Delta_S$ and E$_b$ now depend only on intermonomer coupling and there is one-to-one correspondence
betweeen them. They are, however, not equal, as is sometimes assumed \cite{Feng16a}. This is because between two free triplets there can be no CT
by definition, while in the $^5$(TT)$_1$ of any coupled species in which the individual units are larger than two-level there is always some CT
involving nondegenerate MOs, as is indicated in Fig.~4. E$_b$ is therefore slightly larger than $\Delta_S$, as found is in Table 2.   
We see that our calculated quantities in BP1 and $p-2$ are close to one another. While structural relaxation effects have been ignored in our calculations, we note
that in both T$_1$ and $^1$(TT)$_1$ the triplet wavefunctions occupy individual monomers and the contributions of structural relaxations to 
$\Delta_S$ and E$_b$ will likely cancel, at least partially, in the respective energy differences.
E$_b$ in both BP1 and $p-2$ likely exceeds thermal energy, explaining the long lifetime of the $^1$(TT)$_1$.

Assumption of frozen spin configurations on the alternant (bipartite) phenylene linker suggests ferromagnetic spin-spin
correlation between substituents at meta positions, and negative $\Delta_S$ and E$_b$ in $m-2$ \cite{Abraham17a}.
From our many-body calculations we find both to be positive, albeit very small. This weak deviation from the prediction in Reference \onlinecite{Abraham17a}
can be explained within valence bond (VB) theory, as indicated in Fig.~5.
The meta linkage can be described by {\it spin singlet} VB diagram with ``crossing'' bonds,  which is a superposition of the more familiar Kekul\'e and Dewar VB diagrams. 
Weak but nonzero charge-transfer will occur across the spin-singlet bond between the monomers even with meta-linkage, lowering the 
energy of $^1$(TT)$_1$ relative to $^5$(TT)$_1$ very slightly, and also making E$_b$ positive. Inclusion of realistic second-neighbor electron hopping in Eq.~1 will further
enhance $\Delta_S$ and E$_b$.
\vskip 1pc
\begin{figure}[H]
\centering
\includegraphics[width=3.in]{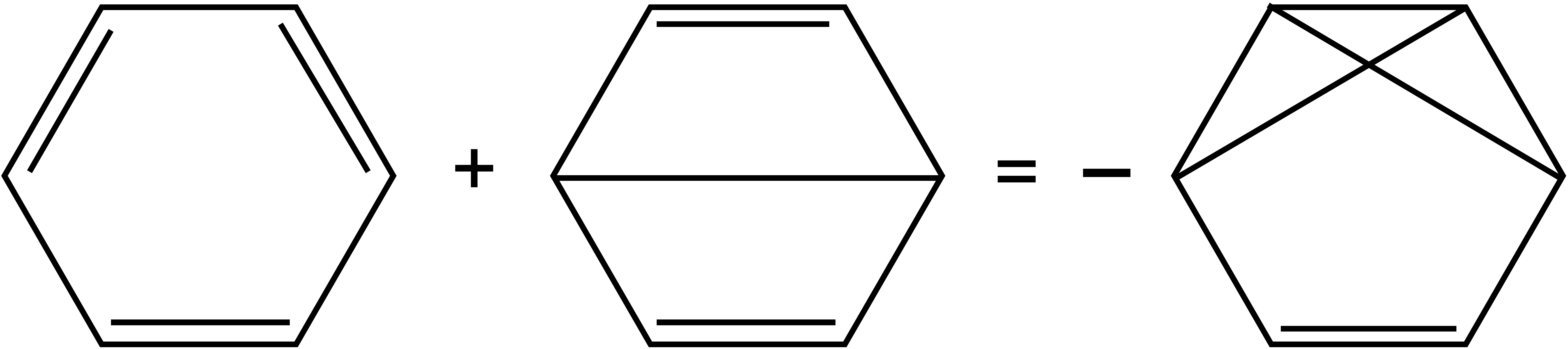}
\caption {\it Linear relationship between VB diagram with
crossed bonds and Kekul\'e and Dewar VB diagrams in benzene.} 
\end{figure}
\vskip 1pc

\section{Conclusions} 
   
To conclude, for moderate to strong intermonomer coupling in ISF compounds, transient absorption measurements
in the IR are essential for distinguishing between free triplets and the bound triplet-triplet in
ISF compounds. In such cases, both the free triplet and the triplet-triplet ESAs in the covalently linked dimers
are different from the triplet absorption in the monomer,
while also being different from one another (see Figs.~3(a) and (c)). This conclusion is independent of the detailed geometry, and is valid for both
longitudinal (2-2$^\prime$) and transverse (6-6$^\prime$) coupling between acene monomers. Comparison with experimental ultrafast spectroscopy in $p-2$
leads to the conclusion that the $^1$(TT)$_1$ here has a very long lifetime. Since a long lifetime opens up various channels of
$^1$(TT)$_1$ decay, our calculations provide a diagnostic tool to experimentalists to preselect ISF compounds for application to solar cells.

In contrast to $p-2$ we find that the intermonomer coupling in $m-2$ is extremely weak, to the extent that the triplet wavefunctions here 
occur as degenerate pairs, with the lowest two consisting of the Frenkel excitons localized on one or the other TIPS-pentacene monomer. 
Thus topology can indeed play a very strong role in ISF.
We agree with 
the authors of reference \onlinecite{Abraham17a} that $\Delta_S$ and E$_b$ are tiny, except that we find these to be still weakly positive. Going beyond the
frozen spin configuration and proper consideration of
electron correlation effects is essential to arrive at the correct state ordering within the triplet-triplet manifold. 
Spontaneous generation of $^5$(TT)$_1$ as well as of free triplets in $m-2$, due to thermal effects or structural relaxations not taken into consideration
in our calculations, are  both possible. We believe that similar tiny energy differences also characterizes BP3, where $^5$(TT)$_1$ has been detected and characterized 
\cite{Tayebjee17a}. Neither the experimental nor the computational free triplet and triplet-triplet ESA spectra are distinguishable in $m-2$. 
It is  therefore conceivable, even likely that free triplets are indeed generated in $m-2$, as claimed in reference \onlinecite{Zirzlmeier15a}.
In agreement with
our conclusion, it has been found that in meta-linked BP1, photoexcitation leads to significant free triplet population lasting into $\mu$s, in contrast to the ``usual''
para-linked BP1, where there is little free triplet generation (private communication, M. Sfeir). Our conclusions 
regarding free triplet generation are slightly
different from those in reference \onlinecite{Korovina18a}, which investigated tetracene dimers and concluded that free triplets are generated from
the para but not the meta-isomer. It is conceivable that the difference, particularly in the case of the meta compound, arises from the $^1$(TT)$_1$ in the
tetracene dimer occurring above S$_1$
(this would explain the fast radiative relaxation here). Ultrafast spectroscopy here was carried out only in the visible wavelength range. Extending these measurements
to the IR should provide additional valuable information.

Two other observations are worthy of noting. First, our calculations indicate that not only $^1$(TT)$_1$ ESA, but even the ground state absorption and the free triplet ESA
provide information on the strength of the intermonomer coupling. Second, the same intermonomer electronic coupling that presumably drives a fast S$_1$ to $^1$(TT)$_1$
internal conversion slows down the $^1$(TT)$_1$ dissociation. For efficient application of SF, this conundrum has to be resolved. 




\bibliography{proposal-chem-new.bib}

\end{document}